%% file: MRC_arxiv.tex
\documentclass[conference]{IEEEtran}
\IEEEoverridecommandlockouts
\usepackage{cite}
\usepackage{amsmath,amssymb,amsfonts}
\usepackage{algorithmic}
\usepackage{textcomp}
\usepackage{xcolor}
\usepackage{cuted}
\def\BibTeX{{\rm B\kern-.05em{\sc i\kern-.025em b}\kern-.08em
    T\kern-.1667em\lower.7ex\hbox{E}\kern-.125emX}}

\usepackage[acronyms,nonumberlist,nopostdot,nomain,nogroupskip]{glossaries}
 
\usepackage{subcaption}

\usepackage{tikz}
\usepackage{tikz-3dplot}
\usepackage{pgfplots}
\usepackage{xcolor}

\def \fwidth{0.8\linewidth}
\def \fheight{0.45\linewidth}

\colorlet{mycolor1}{cyan}
\colorlet{mycolor2}{orange}
\colorlet{mycolor3}{violet}
\colorlet{mycolor4}{red}
\colorlet{mycolor5}{blue}
   
\include{Glossary}

\begin{document}

\title{Geometry and Wideband Performance of a Maximal Ratio Combining Beam}

\author{\IEEEauthorblockN{Andrea Bedin}
\IEEEauthorblockA{\textit{Department of Information Engineering} \\
\textit{University of Padova}\\
Padova, Italy}
\thanks{ This work has received funding from the European Union's
EU Framework Programme for Research and Innovation Horizon 2020 under
Grant Agreement No 861222.}
\IEEEauthorblockA{\textit{Nokia Bell Labs}\\
Espoo, Finalnd\\
andrea.bedin.2@phd.unipd.it} \and
\IEEEauthorblockN{Andrea Zanella}
\IEEEauthorblockA{\textit{Department of Information Engineering} \\
\textit{University of Padova}\\
Padova, Italy\\
andrea.zanella@unipd.it\\}}

\maketitle

\begin{abstract}
This paper discusses the geometrical features and wideband performance of the beam with maximal ratio combining coefficients for a generic multi-antenna receiver. In particular, in case the channel is a linear combination of plane waves, we show that such a beam can be decomposed in a linear combination of beams pointed in the direction of each plane wave, and we compute how many directions can be effectively utilized. This highlights that such beam is better exploiting the spatial diversity provided by the channel, and therefore it is expected to be more robust to disruptions. Moreover, we compute the achieved Signal-to-Noise-Ratio for a wideband receiver, showing that it is not significantly worse than for other methods. Finally, we provide some insights on the robustness of the method by simulating the impact of the blockage of one multipath components.
\end{abstract}
\glsresetall


\section{Introduction}
Modern communication systems often use a codebook-based beamforming approach, where the beams in the codebook are concentrating the gain on a single direction. This approach, while providing good average data rate and implementation simplicity, suffers from the lack of spatial diversity, as it commits on using almost exclusively the multipath component in the high-gain direction. This results in large \gls{snr} drops when the selected component is disrupted by, e.g., a blocker. In contrast, \gls{mrc} is a well-known technique to combine signals received by a multi-antenna system, dating back to 1954 \cite{MRC}. Despite being so dated, it is still widely used and it has proven to be robust and to provide good performance in all sorts of communication conditions.  The classical derivation of \gls{mrc} comes from the \gls{snr} maximization problem in narrowband scenarios \cite{goldsmith}.  Modern communication systems, however, are typically wideband and use analog beamforming, and are thus not capable of fully exploiting the linear gain of \gls{mrc}. Nevertheless, \gls{mrc} turns out to be robust and effective also in such systems, and understanding the reason for this unexpectedly good performance is not straightforward. \\
In this paper, we investigate this aspect by analyzing the geometric features of the \gls{af} of the beam with \gls{mrc} coefficients. Moreover, we analyze the performance of \gls{mrc} outside its design coherence bandwidth, evaluating the \gls{snr} that can be obtained by an analog beamforming wideband system with such a beam.  
Although over the years many analysis \cite{MRC_Weibull, MRC_csi_noise, MRC_nakagami} and variants \cite{MRC_SC, MRC_sequential} of \gls{mrc} have been proposed, to the best of our knowledge, these characteristics of the method have never beed investigated. The main contributions of this work can be hence summarized as follows:
\begin{itemize}
\item We show that, when the channel is a linear combination of plane waves, the beam with the \gls{mrc} coefficients can be decomposed in a linear combination of beams, each pointed towards one of the plane waves;
\item We provide a statistical characterization of the number of beam components that are actually active (i.e., are weighted with a relevant coefficient in the linear combination);
\item We compute the average \gls{snr} achieved by the beam in a wideband setting, and compare it with that achieved by a single-direction beam pointed in the best direction;
\item We provide a numerical evaluation of the distribution of the achieved \gls{snr} when one component of the channel is blocked, and compare it to the single-direction beam solution, to demonstrate the robustness of \gls{mrc}.
\end{itemize}
These results highlight how such beam is better suited for \gls{urllc} than the single-direction beam approach, as it inherently provides more diversity. Hence, considering the importance of \gls{urllc} in the modern communication scene \cite{urllc, urllc2, urllc3}, this beamforming technique should be considered as an alternative to classical beamforming methods.
Furthermore, we observe that implementing such beam is feasible in practice. For example, the \gls{csi} for the beam design can be acquired with a low-cost low-bandwidth digital beamforming chain working alongside the wideband analog beamforming \cite{arch_patent}, or with other methods such as using reference tones \cite{tone_BF}. Moreover, the design of the \gls{mrc} beam has negligible computational complexity, as it only involves the computation of the complex conjugate of the channel coefficients.

\section{System model}

In this paper, we consider a system with an arbitrary antenna array with antennas in positions $A_1$ to $A_N$. When a plane wave is impinging on the array in direction $\vec{k}$, the projection of $A_n$ on $\textit{span}\left\{\vec{k}\right\}$ is denoted by$P_{n,\vec{k}}$, and is given by
\begin{equation}
P_{n, \vec{k}} = \frac{A_n^{T} \vec{k}}{ \vec{k}^{T} \vec{k}} \vec{k}.
\end{equation}

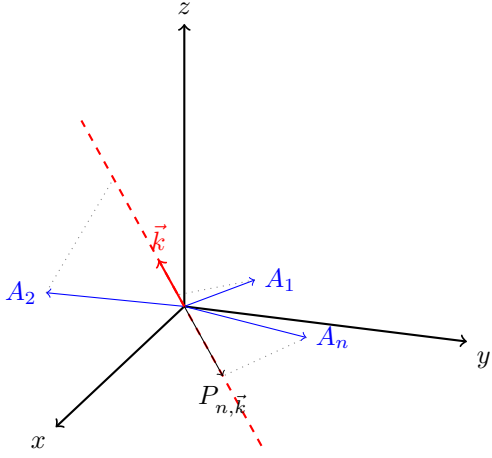
\begin{figure}[t]
\centering
\input{figures/array}
\caption{Array model.}
\label{fig:array}
\end{figure}

In Fig.~\ref{fig:array} we can see that the distance between the projection point and the origin equals the distance traveled by the wave before reaching antenna $n$.
Assuming phase $0$ at the origin, and calling $\lambda$ the wavelength of the signal, this means that the phase observed by antenna $n$ is 
\begin{equation}
\phi_{n, \vec{k}} = \frac{\vert P_{n, \vec{k}} \vert}{\lambda} = \frac{A_n^{T} \vec{k}}{ ( \vec{k}^{T} \vec{k} ) \lambda}  \vert \vec{k} \vert = \frac{A_n^{T} \vec{k}}{  \vert \vec{k} \vert \lambda}.
\end{equation}
Therefore, if we assume $M$ multipath components with amplitude $\{\alpha_m\}$, direction $\{\vec{k}_m\}$ and delay $\{\tau_{m}\}$, the channel frequency response (neglecting the frequency dependence of $\phi_{n, \vec{k}_m}$) at antenna $n$ is  
\begin{equation}
H_n(f) = \sum_{m=1}^{M} \alpha_m e^{j \phi_{n, \vec{k}_m}} e^{-j 2 \pi f \tau_{m}}. \label{eq:chmod}
\end{equation}

We assume analog beamforming is performed according to \gls{mrc} for the center of the frequency band, i.e., the beamforming coefficient for antenna $n$ is\footnote{This could be normalized to have a unitary beamforming vector, however this normalization does not impact the conclusion of this work and is therefore unnecessarily cumbersome.} 
\begin{equation}
\beta_n = \frac{1}{N} H^*_n(0) = \frac{1}{N} \sum_{m=1}^{M} \alpha^*_m e^{-j \phi_{n, \vec{k}_m}}.
\end{equation}

Finally, we assume that the system is affected by Gaussian noise with standard deviation $\sigma_{n_0}$ at each antenna.

\section{Beam Geometry} \label{sec:geometry}

\begin{figure}[t]
\centering
     \begin{subfigure}[t]{0.45\linewidth}
         \includegraphics{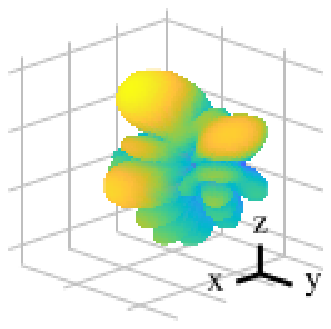}
         \caption{Ineffective component ($\alpha_4 = 0.15$).}
         \label{fig:ineff_example}
     \end{subfigure}
     \begin{subfigure}[t]{0.45\linewidth}       
         \includegraphics{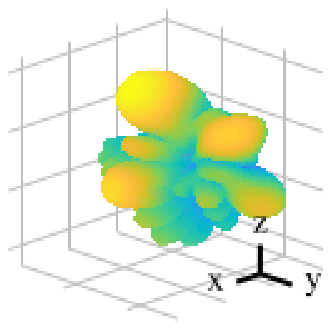}
         \caption{Effective component ($\alpha_4 = 0.6$).}
         \label{fig:eff_example}
     \end{subfigure}
        \caption{Change in radiation pattern induced by an effective and an ineffective component.}
        \label{fig:pattern_eff_example}
\end{figure}

The array factor in direction $\vec{r}$, with $\vert \vec{r} \vert = 1$, is 
\begin{align}
F(\vec{r}) &= \sum_{n=1}^{N} \beta_n e^{j \phi_{n, \vec{r}}}\\ 
&= \frac{1}{N} \sum_{n=1}^{N} \left(  \sum_{m=1}^{M} \alpha^*_m e^{-j \phi_{n, \vec{k}_m}} \right) e^{j \phi_{n, \vec{r}}}.
\end{align}
By rearranging the sums, we can highlight the contribution of each multipath component to the array factor, obtaining the expression:
\begin{equation}
F(\vec{r}) =  \sum_{m=1}^{M}  \alpha^*_m \frac{1}{N} \left( \sum_{n=1}^{N} e^{-j \phi_{n, \vec{k}_m}}  e^{j \phi_{n, \vec{r}}} \right) = \sum_{m=1}^{M} \alpha^*_m F_m(\vec{r})
\end{equation}
where 
\begin{equation}
F_m(\vec{r}) = \frac{1}{N} \sum_{n=1}^{N} e^{-j \phi_{n, \vec{k}_m}}  e^{j \phi_{n, \vec{r}}},
\end{equation}
which denotes the array factor component associated to the multipath component $m$.
Clearly, it holds $F_m(\vec{r}) \leq 1$ and 
\begin{equation}
F_m(\vec{k}_m) = \frac{1}{N} \sum_{n=1}^{N} e^{-j \phi_{n, \vec{k}_m}}  e^{j \phi_{n, \vec{k}_m}}  = \frac{1}{N} \sum_{n=1}^{N} 1 = 1.
\end{equation}
Therefore, we can conclude that each array factor component has a global maximum in the direction of the plane it is associated with. Let us now determine the gain observed by a generic component. For the generic direction $\vec{k}_h$, we obtain a total gain:
\begin{equation}
F(\vec{k}_h) =  \alpha^*_h + \frac{1}{N} \sum_{\substack{m=1 \\ m\neq h}}^{M}  \alpha^*_m \left( \sum_{n=1}^{N} e^{-j \phi_{n, \vec{k}_m}}  e^{j \phi_{n, \vec{k}_h}} \right). \label{eq:decomposition_gain}
\end{equation}
If the multipath components are few and spread apart, and the array factor components are narrow beams, i.e., the gain rapidly decreases moving away form the maximum, we have that the second term of the sum is small, therefore $F(\vec{k}_h) \approx  \alpha^*_h$. In other words, the \gls{mrc} between the antennas is equivalent to the \gls{mrc} between the components. On the other hand, if the amplitude of the $h$-th component is small, and a lot of other components are present, the second term in \eqref{eq:decomposition_gain} becomes relevant. In this case, the $h$-th component might bring negligible contribution to the total received power. For this reason, we define a condition of effectiveness for the component according to which component $h$ is effective if
\begin{equation}
\vert \alpha^*_h \vert \geq \vert X_h \vert, \label{eq:effcondition}
\end{equation}
where 
\begin{equation}
X_h = \sum_{\substack{m=1 \\ m\neq h}}^{M}  \frac{1}{N} \alpha^*_m \left( \sum_{n=1}^{N} e^{-j \phi_{n, \vec{k}_m}}  e^{j \phi_{n, \vec{r}_1}} \right).
\label{eq:X_def}
\end{equation}
To understand this definition, let us consider the impact of amplitude variations of the $h$-th multipath component on the overall channel. In case the component is effective, we have that 
\begin{equation}
\frac{\partial  H(f) }{\partial  \alpha_h } = \frac{\partial F(\vec{k}_h) \alpha_h}{\partial \alpha_h} \approx \frac{ \partial \vert \alpha_h \vert^2}{\partial \alpha_h}.
\end{equation}
In contrast, in the ineffective case we have 
\begin{equation}
\frac{\partial  H(f) }{\partial  \alpha_h } = \frac{\partial F(\vec{k}_h) \alpha_h}{\partial \alpha_h} \approx  \frac{\partial X_h \alpha_h }{\partial \alpha_h}.
\end{equation}
Clearly, this shows how changing the amplitude of an effective components has a quadratic effect on the channel, whereas an ineffective component will have only a linear impact. Moreover, the impact of an ineffective component is scaled by $X_h$, that when the component is ineffective, is by definition smaller than the amplitude of the effective components. Therefore, a disruption of an ineffective component will affect the channel negligibly. To visually exemplify the definition, in Fig.~\ref{fig:pattern_eff_example} we show the array pattern for a channel with the parameters listed in Tab.~\ref{tab:example_chparams}. In this case, $X_4 = 0.32$ and we plot the pattern for $\alpha_4 = 0.15$ (Fig.~\ref{fig:ineff_example}) and $\alpha_4 = 0.6$ (Fig.~\ref{fig:eff_example}). It can be clearly seen that, in the effective case, the beam pattern has an additional lobe in the direction of $k_4$.

\begin{table}[t]
\caption{Example channel parameters}
\begin{center}
\begin{tabular}{c|c|c}
$k$ & $\vec{k}_m$ & $\alpha_m$ \\
\hline
$1$ & $(-1,0,0)$ & $0.5$ \\
$2$ & $\left( \frac{1}{\sqrt{3}},-\frac{1}{\sqrt{3}},-\frac{1}{\sqrt{3}} \right)$ & $1$ \\
$3$ & $\left( -\frac{1}{\sqrt{2}},0,\frac{1}{\sqrt{2}} \right)$ & $1.5$ \\
$3$ & $\left(-\frac{1}{\sqrt{3}}, -\sqrt{\frac{2}{3}},0 \right)$ & $\alpha_4$ 
\end{tabular}
\label{tab:example_chparams}
\end{center}
\end{table}

%\begin{figure}[t]
%\centering
%     \begin{subfigure}[t]{0.45\linewidth}
%         \input{figures/3dpattern_small}
%         \caption{Ineffective component ($\alpha_4 = 0.15$).}
%         \label{fig:ineff_example}
%     \end{subfigure}
%     \begin{subfigure}[t]{0.45\linewidth}       
%         \input{figures/3dpattern_large}
%         \caption{Effective component ($\alpha_4 = 0.6$).}
%         \label{fig:eff_example}
%     \end{subfigure}
%        \caption{Change in radiation pattern induced by an effective and an ineffective component.}
%        \label{fig:pattern_eff_example}
%\end{figure}

To characterize the probability of effectiveness of the components, we consider the following assumptions: 
\begin{enumerate}
\item $\alpha_m$ are distributed according to a complex normal \gls{rv}, $\mathcal{CN}(0, 1)$;
\item $\vec{k}_m$ can have an arbitrary distribution, however they are typically taken as uniformly distributed within  elements aperture;
\item all $\alpha_m$ and $\vec{k}_m$  are mutually uncorrelated.
\end{enumerate}
With these assumptions, we study the random variable
\begin{equation}
X = X_1 = \sum_{m=2}^{M}  \frac{1}{N} \alpha^*_m \left( \sum_{n=1}^{N} e^{-j \phi_{n, \vec{k}_m}}  e^{j \phi_{n, \vec{k}_1}} \right).
\label{eq:X_def}
\end{equation}
where, without loss of generality, we consider $h=1$.
Clearly, as each component of the sum is zero mean, we have
\begin{equation}
\mathbb{E} \left[ X \right] = 0 .
\end{equation}
Since $\{\alpha_m\}$ has unitary variance by assumption, and amplitudes and angles are independent, by defining $\bar{M} = M-1$ we have:
\begin{align}
Var \left[ X \right]& = \mathbb{E}  \left[ \left\vert  \sum_{m=2}^{M}  \alpha^*_m \left(  \frac{1}{N}\sum_{n=1}^{N} e^{-j \phi_{n, \vec{k}_m}}  e^{j \phi_{n, \vec{r}}} \right)\right\vert^2 \right] \label{eq:expexted_sidelobe_first}\\
&= \mathbb{E}  \left[  \sum_{m=2}^{M} \left\vert \alpha_m \left(  \frac{1}{N}\sum_{n=1}^{N} e^{-j \phi_{n, \vec{k}_m}}  e^{j \phi_{n, \vec{k}_1}} \right) \right\vert^2  \right] \\
 &= \bar{M} \mathbb{E}  \left[ \left\vert \left(  \frac{1}{N}\sum_{n=1}^{N} e^{-j \phi_{n, \vec{k}_m}}  e^{j \phi_{n, \vec{k}_1}} \right) \right\vert^2  \right]\\
&=  \bar{M} a^2; \label{eq:expected_sidelobe}
\end{align}
where 
\begin{equation}
a = \mathbb{E}  \left[ \left\vert \left(  \frac{1}{N}\sum_{n=1}^{N} e^{-j \phi_{n, \vec{k}_m}}  e^{j \phi_{n, \vec{k}_1}} \right) \right\vert^2  \right] \label{eq:a_def}
\end{equation}
is the only parameter that depends on the array geometry, and therefore we call it the \textit{array parameter}. This value can be computed numerically for the array of interest through, e.g., Monte Carlo simulation. 
We can then approximate $X$ by a zero mean complex Gaussian \gls{rv} with variance $\bar{M}a^2$. Such approximation is suggested by the law of large numbers, but it is not necessarily verified in practice. Nonetheless, this approximation is mathematically convenient, and it determines a relativelysmall gap with physically-accurate simulations, as we will show in the results section. With this approximation, the conditional probability of ineffectiveness given $\alpha_1$ is the probability that $\vert \alpha_1 \vert$ is smaller than $\vert X \vert$, which is a Rayleigh \gls{rv} of parameter $\sqrt{\frac{\bar{M}}{2}}a$. Therefore we have
\begin{equation}
P_{ineff}( z ) \triangleq P \left[\vert X \vert \geq \vert \alpha_1 \vert \:\bigg\vert \: \vert \alpha_1 \vert = z \right]   =  e^{-\frac{z^2}{\bar{M} a^2}}. 
\label{eq:approx_p_of_alpha}
\end{equation}  

The overall $P_{ineff}$ can be obtained by integrating $P_{ineff}(z)$ over the distribution of $\vert \alpha_1 \vert$, which is Rayleigh with parameter $\frac{1}{\sqrt{2}}$. This can be expressed as
\begin{align}
P_{ineff} &=   \int_{0}^{\infty} P\left[ \vert \alpha_1  \vert = z \right] P_{ineff}(z) d z\\
&=  \int_{0}^{\infty}  2 z e^{-z^2}   e^{-\frac{z^2}{ \bar{M} a^2}}  dz\\
&= \frac{\bar{M} a^2}{1 + \bar{M}a^2}. \label{eq:pineff_gauss}
\end{align}

\section{Wideband behavior} \label{sec:wideband}

\subsection{\gls{mrc} performance}
When the \gls{mrc} beam is used in an analog beamforming wideband system, the phases of the channel coefficients are frequency dependent, therefore the classical \gls{snr} formulation does not apply outside one coherence bandwidth from the carrier frequency. Instead, the received power outside of the coherence bandwidth can be expressed as

\begin{align}
H(f) =& \sum_{m = 1}^{M} \vert \alpha_m \vert^2 e^{-j 2 \pi f \tau_{m}} + \sum_{m = 1}^{M} \alpha_m e^{-j 2 \pi f (\tau_{m} - \tau_{m'})} \nonumber \\
& \sum_{m' \neq m}  \frac{1}{N} \alpha^*_{m'} \left( \sum_{n=1}^{N} e^{-j \phi_{n, \vec{k}_{m'}}}  e^{j \phi_{n, \vec{k}_m}}  \right), \label{eq:chresp}
\end{align} 

where the first summation accounts for the contribution of the $M$ multipath components received by the corresponding beam components, while the other term is the aggregate contribution of the multipath components not aligned with the beam components.
Considering a frequency well outside the coherence bandwidth of the channel, we can assume that the phases $2 \pi f(\tau_{m} - \tau_{m'})$ are uniformly distributed and independent. With this assumption, the expected channel power gain is
\begin{align}
&\mathbb{E}\left[ \vert H(f) \vert^2 \right] = \sum_{m = 1}^{M} \mathbb{E}\left[ \vert \alpha_m \vert^4 \right] \label{eq:gain_full} \\&+   \sum_{m = 1}^{M} \mathbb{E}\Biggl[ \vert \alpha_m \vert^2 \left\vert \sum_{m' \neq m}  \frac{1}{N} \alpha^*_{m'} \left( \sum_{n=1}^{N} e^{-j \phi_{n, \vec{k}_{m'}}}  e^{j \phi_{n, \vec{k}_m}}  \right)  \right\vert^2 \Biggl]. \nonumber  
\end{align}

Recalling the independence between paths and the definition of the \textit{array parameter} $a$ in \eqref{eq:a_def}, we can rewrite \eqref{eq:gain_full} as
\begin{equation}
\mathbb{E}\left[ \vert H(f) \vert^2 \right] = M \left(  \mathbb{E} \left[\vert \alpha_m \vert^4 \right]  +  \mathbb{E} \left[\vert \alpha_m \vert^2 \right] \bar{M}a^2 \right).
\end{equation}

Using the assumption that $\alpha_m \sim \mathcal{CN}(0,1)$, the expectations $\mathbb{E} \left[\vert \alpha_m \vert^4 \right]$ and $\mathbb{E} \left[\vert \alpha_m \vert^2 \right]$ are the $4^\textit{th}$ and $2^\textit{nd}$ moment of a Rayleigh \gls{rv} with parameter $\frac{1}{\sqrt{2}}$, which are $2$ and $1$, respectively. Thus, the final expression for the gain is 
\begin{equation}
\mathbb{E}\left[ \vert H(f) \vert^2 \right] = M \left(  2  +   \bar{M}a^2 \right).
\end{equation}

The noise is a linear combination of Gaussian \glspl{rv} with coefficients $\beta_n$, therefore the variance is
\begin{equation}
\sigma^2_n = \sigma_{n_0}^2 \sum_{n=1}^{N}  \vert \beta_n \vert^2,
\end{equation}
with expected value  
\begin{equation}
\mathbb{E} \left[ \sigma^2_n \right] = N \sigma_{n_0}^2 \mathbb{E} \left[ \vert \beta_n^2 \vert \right]. \label{eq:noise_expect}
\end{equation}
The right-most expectation in \eqref{eq:noise_expect} can be computed as
\begin{align}
\mathbb{E}\left[ \vert \beta_n \vert^2 \right] &= \mathbb{E} \left[ \left\vert \frac{1}{N} \sum_{m=1}^{M} \alpha^*_m e^{-j \phi_{n, \vec{k}_m}} \right\vert^2 \right]\\ 
&=\frac{1}{N^2} \sum_{m=1}^M \mathbb{E} \left[ \vert \alpha_m \vert^2 \right] = \frac{M}{N^2},
\end{align}

so that the expected noise variance is
\begin{equation}
\mathbb{E} \left[ \sigma^2_n \right] = N \sigma_{n_0}^2 \frac{M}{N^2} = \frac{M}{N} \sigma_{n_0}^2.
\end{equation}

Based on these results, we can compute the average-signal-to-average-noise-ratio as
\begin{align}
\Gamma = \frac{\mathbb{E}\left[\vert H(f)\vert^2 \right]}{\mathbb{E}\left[\sigma_n^2 \right]} &= M \left(  2  +   \bar{M}a^2 \right) \frac{N}{M \sigma_{n_0}^2}\\
&=  \frac{N}{\sigma_{n_0}^2} \left(  2  +   \bar{M}a^2 \right). \label{eq:snr_mrc}
\end{align}
Note that, despite the function looks linear in $N$, this is not guaranteed. In fact, the \textit{array parameter} $a$ also depends in a complicated manner on the array size and geometry. Moreover, for $M=1$, \eqref{eq:snr_mrc} gives a $2N$ gain compared to the \gls{snr} observed by a single element. This is actually an artifact of approximating the \gls{snr} using the average signal to average noise ratio. In this case in fact, the \gls{mrc} corresponds to the classical beamforming, which is known to have an \gls{snr} gain of $N$. We therefore note that the proposed approximation has a $3$dB error for $M=1$.

\begin{table}[t]
\caption{\textit{Array parameters} for some \glspl{ula}}
\begin{center}
\begin{tabular}{c|c|c|c}
\gls{fov} & $180^\circ$ & $120^\circ$ & $60^\circ$ \\
\hline
\textbf{\gls{ula} Elements} & \multicolumn{3}{|c}{$a$}  \\
\hline
2 & 0.55 & 0.50 & 0.69\\
4 & 0.30 & 0.26 & 0.40\\
8 & 0.17 & 0.14 & 0.22\\
16 & 0.09 & 0.07 & 0.12\\
32 & 0.05 & 0.04 & 0.06\\
64 & 0.03 & 0.02 & 0.03\\
\hline
\end{tabular}
\label{tab:sigmasULA}
\end{center}
\end{table}

\begin{figure}[t]
     \centering
     \begin{subfigure}[b]{0.95\linewidth}
         \input{figures/pineff}
         \caption{Ineffectiveness probability.}
         \label{fig:pineff}
         \vspace{5mm}
     \end{subfigure}
     \begin{subfigure}[b]{0.95\linewidth}       
         \input{figures/ncomp}
         \caption{effective components.}
         \label{fig:ncomp}
     \end{subfigure}
        \caption{(a) ineffectiveness probability and (b) number of components utilized obtained by \gls{mrc} for different \glspl{ula} (\gls{fov} $180^\circ$). The lines represent the theoretical values, whereas the marks are given by numerical evaluation.}
        \vspace{-5mm}
        \label{fig:components}
\end{figure}
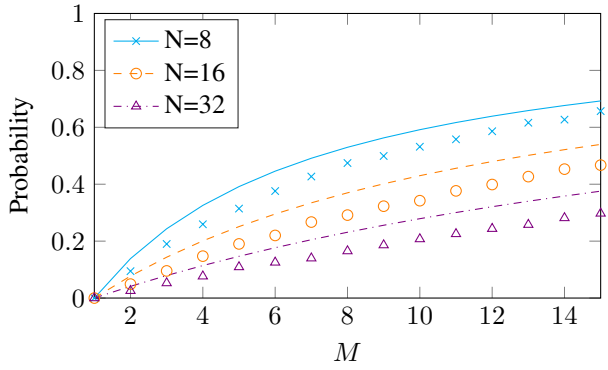
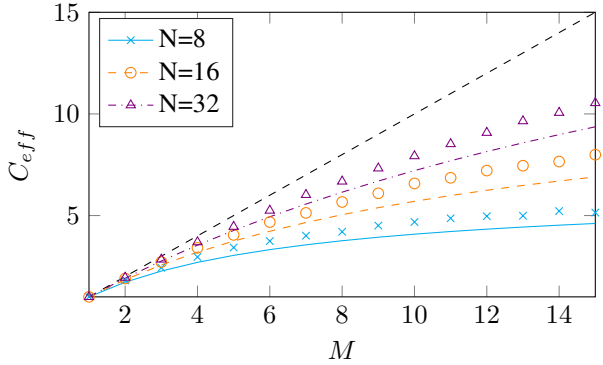

\begin{figure*}[!ht]
     \centering
          \begin{subfigure}[t]{0.9\textwidth}
         \centering
         \input{figures/legend_SNR}
         \label{fig:legend}
     \end{subfigure}
     \begin{subfigure}[b]{0.32\linewidth}
         \input{figures/SNR_8ant_180deg}
         \caption{$N=8$ .}
         \label{fig:SNR_8ant}
     \end{subfigure}
     \begin{subfigure}[b]{0.32\linewidth}       
         \input{figures/SNR_16ant_180deg}
         \caption{$N=16$ .}
         \label{fig:SNR_16ant}
     \end{subfigure}
          \begin{subfigure}[b]{0.32\linewidth}       
         \input{figures/SNR_32ant_180deg}
         \caption{$N=32$.}
         \label{fig:SNR_32ant}
     \end{subfigure}
        \caption{SNR obtained by \gls{mrc} and the single-direction beam for different \glspl{ula}, for a \gls{fov} of $180^\circ$ and $\sigma_{n_0}=1$, and for different values of the number $N$ of antennas. }
        \label{fig:SNRs}
\end{figure*}
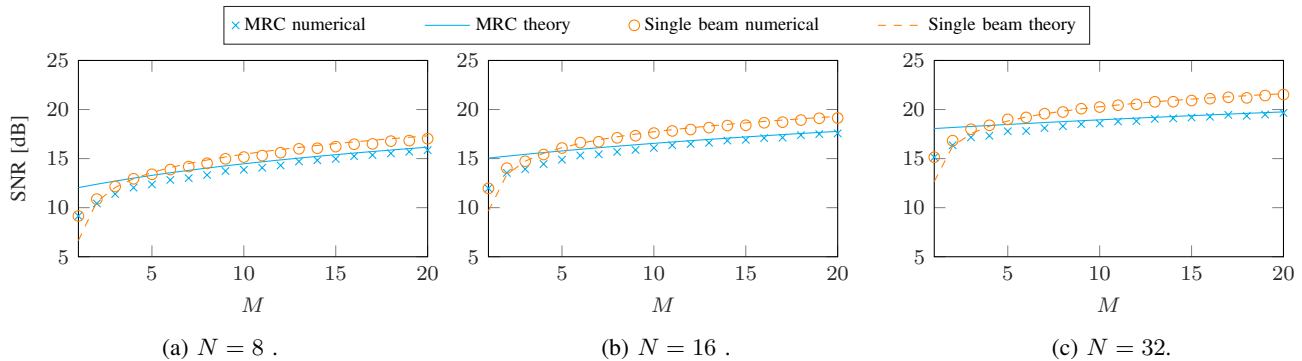

\begin{figure}[!ht]
\centering
\input{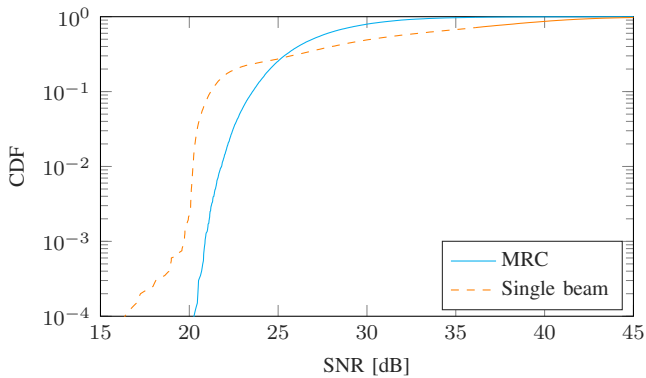}
\caption{\gls{snr} distribution for an 8 element \gls{ula} with a \gls{fov} of $180^\circ$, $M=20$  and $\sigma_{n_0}=1$.}
\label{fig:snr_cdf}
\end{figure}

\subsection{Single-direction beam performance}
As a comparison, we compute the gain obtained with a single beam pointed towards the largest component that, without loss of generality, we assume to be the first. Therefore, we set the beam coefficients to
\begin{equation}
\beta^{(Sing)}_n = \frac{1}{N} e^{-j \phi_{n, \vec{k}_1}}.
\end{equation}
With this assumption, and again using the definition of \textit{array parameter}, the channel power gain can be written as
\begin{equation}
\mathbb{E} \left[ \vert H^{(Sing)}(f) \vert^2 \right] =  \mathbb{E} \left[ \vert \alpha_1 \vert^2 \right] + \sum_{m=2}^M a^2 \mathbb{E} \left[ \vert \alpha_m \vert^2 \right] .
\end{equation}
Note that, under the assumption that $\alpha_1$ is the largest component, its statistical distribution changes. In fact, if we assume that a generic $\alpha_m$ has exponentially distributed power with parameter 1 (which is a direct consequence of the Gaussian distribution of $\alpha_m$), the \gls{cdf} of $\vert \alpha_1 \vert^2$ is given by 
\begin{equation}
P \left[ \vert \alpha_1 \vert^2  < x \right] = \left( 1-e^{-x}\right)^M,
\end{equation}
and its \gls{pdf} is hence
\begin{equation}
\frac{\partial}{\partial x} \left( 1-e^{-x}\right)^M = M e^{-x} \left(1-e^{-x}\right)^{M-1}.
\end{equation}
Finally, its expected value is 
\begin{equation}
\int_0^{\infty } M x e^{-x} \left(1-e^{-x}\right)^{M-1}  dx = H_M \approx \log(M) + \gamma,
\end{equation}
where $H_M$ is the $M$-th harmonic number and $\gamma$ is the Euler's constant. 
The statistics of $\alpha_m$ for $m \neq 1$ would also change, but we neglect this aspect, thus obtaining 
\begin{equation}
\mathbb{E} \left[ \vert H^{(Sing)}(f) \vert^2 \right] \approx  \log(M) +\gamma + \bar{M} a^2.
\end{equation}
The expected noise power will be simply 
\begin{equation}
\mathbb{E} \left[ \sigma^2_n \right] = \frac{\sigma_{n_0}^2}{N}, 
\end{equation}
and the \gls{snr} is 
\begin{equation}
\Gamma^{(Sing)} = \frac{\mathbb{E} \left[ \vert H(f) \vert^2 \right]}{\mathbb{E} \left[ \sigma^2_n \right]} = \frac{N (\log(M) + \gamma + \bar{M} a^2) }{\sigma^2_{n_0}}. \label{eq:snr_sing}
\end{equation}
The ratio between the \gls{snr} with a single-direction and with \gls{mrc} is hence 
\begin{align}
\frac{\Gamma^{(Sing)}}{\Gamma} &= \frac{N (\log(M) + \gamma + \bar{M} a^2) }{\sigma^2_{n_0}} \frac{\sigma_{n_0}^2}{N \left( 2  +  \bar{M}a^2 \right) } \\
&= \frac{ (\log(M) + \gamma + \bar{M} a^2) }{ \left( 2  +  \bar{M}a^2 \right) }. \label{eq:ratio_SNRS}
\end{align}

We note that
\begin{equation}
\lim_{M \rightarrow \infty} \frac{ (\log(M) + \gamma + \bar{M} a^2) }{ \left( 2  +  \bar{M}a^2 \right) } = 1, \label{eq:convergence}
\end{equation}
therefore, for rich multipath channels the two methods show the same gain.

\section{Results}

\glsreset{ula}
\glsreset{fov}

In Tab.~\ref{tab:sigmasULA} are listed the values of the \textit{array parameter} $a$ for various \glspl{ula} and uniformly distributed angle of arrival within a fixed \gls{fov}. As expected, the value decreases with the number of antennas, since the probability of the array having large gain in a random direction decreases. It also decreases with the \gls{fov}, as with a smaller \gls{fov} there is less space covered by the sidelobes. \\
To verify the theoretical results, we performed some numerical simulations randomly generating some channels according to \eqref{eq:chmod} and assuming a \gls{ula}. The directions of arrival are uniformly distributed within the \gls{fov} of the array and the delays are uniformly distributed between $0$ and $100$ns. We averaged the channel gain over a bandwidth of $1$GHz.
Fig.~\ref{fig:pineff} shows the probability of ineffectiveness of a component  for different \glspl{ula} as a function of the total number of multipath components. The lines represent the theoretical value according to \eqref{eq:pineff_gauss}, whereas the marks represent the value estimated numerically over 1000 realizations. Similarly, Fig. \ref{fig:ncomp} shows the median number of effective components $C_{eff} = M \left(1- P_{ineff} \right)=  M \left(1- \frac{\bar{M}a^2}{1 + \bar{M}a^2} \right)$. Again, the solid lines represent the theoretical value and the marks represent the numerical estimations. As we can see, in both cases the theory follows the numerical estimation quite closely, with a small gap caused by the gaussian approximation of $X$. We also notice that even with a modest array of only $8$ antennas we can exploit as many as $4$ components in a channel that has a total of only $6$. \\
Fig.~\ref{fig:SNRs} shows how the \gls{snr} changes as a function of $M$ for both the single-direction beam and the \gls{mrc} methods. In particular, the lines are calculated with equations \eqref{eq:snr_mrc} and \eqref{eq:snr_sing} respectively, whereas the marks are the simulated \gls{snr} for the corresponding parameters. Here, it can be clearly seen that the accuracy of the approximation average-signal to average-noise ratio in place of the \gls{snr} improves as $M$ increases. As noted in Sec.~\ref{sec:wideband}, this approximation leads to an error of $3$dB for $M=1$, therefore we expect the error to be always lower than this value. For the proposed configurations, we can observe that the approximation error drops below $1$dB  for $M > 4$. We also note that the gap between the two methods is relatively small, in the order of a few dB. Moreover, despite $M$ is not large enough to show the convergence expected from \eqref{eq:convergence}, such result shows that the gap will not increase in more complex channels. \\
To evaluate the increase of robustness enabled by the diversity generated by the \gls{mrc}, we generated the beam for a given channel $H(f)$, then simulated a blockage by removing component $m'$, where $m'$ is randomly selected between $1$ and $M$, generating the new channel 
\begin{equation}
H'(f) = H(f) - \alpha_{m'} e^{j \phi_{n, \vec{k}_{m'}}} e^{-j 2 \pi f \tau_{m'}}.
\end{equation}
We then applied the beam designed for $H(f)$ to the new channel $H'(f)$ and evaluated the \gls{snr} obtained by the two methods. In Fig.~\ref{fig:snr_cdf} we can observe the resulting \gls{snr} distribution. As expected, we can observe that the tail of the \gls{snr} obtained with a single beam extends much further than that of \gls{mrc}, because the single beam approach is much more sensitive too the loss of the component used by the beam, since the remaining energy only comes from sidelobes.

\section{Conclusions and future work}
In this paper, we studied the properties of the beam generated with \gls{mrc} coefficients, when this is used outside one coherence bandwidth from its design frequency. This provides an evaluation of the performance of such method in a wideband analog beamforming system where the coherence bandwidth is much smaller than the system bandwidth. We have shown that the method can be implemented with minimal degradation of the \gls{snr} compared to the classical beam pointed in a single direction. In contrast, we proved that \gls{mrc} generates a beam with multiple lobes, and therefore it can better exploit the spatial diversity offered  by the environment. Thanks to this additional diversity, it can better handle blockage events, significantly shortening the tail of the distribution of the \gls{snr} when some multipath components are suddenly removed. This allows a trade-off between average rate and robustness, that makes the method a viable choice for ultra reliable communications. As we expect the method to be more susceptible to interference due to the lower selectivity of the beam pattern, in future works we plan to study the interference properties of such beam, and possibly propose interference mitigation techniques to overcome such limitation.

\bibliographystyle{IEEEtran}
\bibliography{refs}

\end{document}

%% file: Glossary.tex
\newacronym{snr}{SNR}{Signal-to-Noise-Ratio}
\newacronym{sir}{SIR}{Signal to Interference Ratio}
\newacronym{af}{AF}{Array Factor}
\newacronym{mrc}{MRC}{Maximal Ratio Combining}
\newacronym{cdf}{CDF}{Cumulatve Distribution Function}
\newacronym{pdf}{PDF}{Probability Density Function}
\newacronym[longplural={Uniform Linear antenna Array}, shortplural={ULAs}]{ula}{ULA}{Uniform Linear antenna Array}
\newacronym{fov}{FoV}{Field of View}
\newacronym[longplural={Random Variables}, shortplural={r.v.s}]{rv}{r.v.}{Random Variable}
\newacronym{urllc}{URLLC}{Ultra Reliable Low Latency Communications}
\newacronym{csi}{CSI}{Channel State Information}

%% file: figures/array.tex
\tdplotsetmaincoords{70}{110}
\begin{tikzpicture}[tdplot_main_coords]
\draw[thick,->] (0,0,0) -- (5,0,0) node[anchor=north east]{$x$};
\draw[thick,->] (0,0,0) -- (0,4,0) node[anchor=north west]{$y$};
\draw[thick,->] (0,0,0) -- (0,0,4) node[anchor=south]{$z$};

\draw[blue, ->] (0,0,0) -- (0,1,0.5) node[anchor=west] {$A_1$};
\draw[blue, ->] (0,0,0) -- (4,-0.5,1.5) node[anchor=east] {$A_2$};
\draw[blue, ->] (0,0,0) -- (-2,1,-1) node[anchor=west] {$A_n$};

\draw[red, ->, thick] (0,0,0) -- (1,0,1) node[anchor=south] {$\vec{k}$};

\draw[red, dashed, thick] (-3,0,-3) -- (4,0,4);

\draw[gray, dotted] (0,1,0.5) -- (0.25,0,0.25);
\draw[gray, dotted] (-2,1,-1) -- (-1.5,0,-1.5);
\draw[gray, dotted] (4,-0.5,1.5) -- (2.75,0,2.75);

\draw[black, ->] (0,0,0) -- (-1.5,0,-1.5) node[anchor=north] {$P_{n,\vec{k}}$};

\end{tikzpicture}

%% file: figures/pineff.tex

%
\begin{tikzpicture}

\begin{axis}[%
width=\fwidth,
height=\fheight,
scale only axis,
unbounded coords=jump,
xmin=1,
xmax=15,
xlabel style={font=\footnotesize\color{white!15!black}},
xlabel={$M$},
xlabel near ticks,
ymin=0,
ymax=1,
ylabel style={font=\footnotesize\color{white!15!black}},
ylabel={Probability},
ylabel near ticks,
axis background/.style={fill=white},
legend style={at={(0.02,0.98)}, anchor=north west, legend cell align=left, align=left, draw=white!15!black}
]
\addplot [color=mycolor1, only marks, mark=x, mark options={solid, mycolor1},forget plot]
  table[row sep=crcr]{%
1	0\\
2	0.0945\\
3	0.19\\
4	0.25925\\
5	0.3144\\
6	0.376\\
7	0.427\\
8	0.4745\\
9	0.499222222222222\\
10	0.5317\\
11	0.557727272727273\\
12	0.586083333333333\\
13	0.616076923076923\\
14	0.627\\
15	0.656933333333333\\
};

\addplot [color=mycolor1, forget plot]
  table[row sep=crcr]{%
1	0\\
2	0.138639706289139\\
3	0.243518130490934\\
4	0.325628922514186\\
5	0.391659959786505\\
6	0.445913348680057\\
7	0.491282163482973\\
8	0.529783647940454\\
9	0.562867325501827\\
10	0.591601590559452\\
11	0.616791246981877\\
12	0.639054075658631\\
13	0.658872177932519\\
14	0.676627285016292\\
15	0.692625586178413\\
};

\addplot [color=mycolor2, only marks, mark=o, mark options={solid, mycolor2},forget plot]
  table[row sep=crcr]{%
1	0\\
2	0.049\\
3	0.0953333333333334\\
4	0.14725\\
5	0.1902\\
6	0.22\\
7	0.267\\
8	0.2915\\
9	0.322777777777778\\
10	0.3421\\
11	0.376636363636364\\
12	0.398916666666667\\
13	0.426692307692308\\
14	0.453\\
15	0.467066666666667\\
};

\addplot [color=mycolor2, dashed, forget plot]
  table[row sep=crcr]{%
1	0\\
2	0.0773367336599728\\
3	0.143570215780615\\
4	0.200931438667614\\
5	0.251091212064512\\
6	0.295325599403655\\
7	0.334625994787078\\
8	0.36977431113331\\
9	0.401395533184458\\
10	0.429995205485918\\
11	0.455986663954789\\
12	0.479711142019143\\
13	0.501452835617012\\
14	0.521450342511285\\
15	0.539905454683801\\
};

\addplot [color=mycolor3, only marks, mark=triangle, mark options={solid, mycolor3},forget plot]
  table[row sep=crcr]{%
1	0\\
2	0.0255\\
3	0.0526666666666666\\
4	0.0765\\
5	0.109\\
6	0.125166666666667\\
7	0.139857142857143\\
8	0.165\\
9	0.186\\
10	0.2073\\
11	0.225272727272727\\
12	0.243583333333333\\
13	0.257846153846154\\
14	0.281357142857143\\
15	0.297333333333333\\
};

\addplot [color=mycolor3, dashdotted, forget plot]
  table[row sep=crcr]{%
1	0\\
2	0.041197742547798\\
3	0.0791352897999713\\
4	0.114184906852673\\
5	0.146664260816899\\
6	0.176846076257939\\
7	0.204965811599835\\
8	0.231227814102802\\
9	0.255810293960698\\
10	0.278869373649545\\
11	0.300542407075465\\
12	0.320950717659222\\
13	0.34020187067001\\
14	0.358391569689127\\
15	0.375605247792908\\
};

\addplot [color=mycolor1, mark=x, mark options={solid, mycolor1}]{
 -1 -1};
\addlegendentry{N=8}

\addplot [color=mycolor2, dashed, mark=o, mark options={solid, mycolor2}]{
 -1 -1};
\addlegendentry{N=16}

\addplot [color=mycolor3, dashdotted, mark=triangle, mark options={solid, mycolor3}]{
 -1 -1};
\addlegendentry{N=32}

\end{axis}

\end{tikzpicture}%

%% file: figures/ncomp.tex

\begin{tikzpicture}

\begin{axis}[%
width=\fwidth,
height=\fheight,
scale only axis,
unbounded coords=jump,
xmin=1,
xmax=15,
xlabel style={font=\footnotesize\color{white!15!black}},
xlabel={$M$},
xlabel near ticks,
ymin=1,
ymax=15,
ylabel style={font=\footnotesize\color{white!15!black}},
ylabel={$C_{eff}$},
ylabel near ticks,
axis background/.style={fill=white},
legend style={at={(0.02,0.98)}, anchor=north west, legend cell align=left, align=left, draw=white!15!black}
]
\addplot [color=black, dashed, forget plot]
  table[row sep=crcr]{%
1	1\\
2	2\\
3	3\\
4	4\\
5	5\\
6	6\\
7	7\\
8	8\\
9	9\\
10	10\\
11	11\\
12	12\\
13	13\\
14	14\\
15	15\\
};

\addplot [color=mycolor1, only marks, mark=x, mark options={solid, mycolor1}, forget plot]
  table[row sep=crcr]{%
1	1\\
2	1.811\\
3	2.43\\
4	2.963\\
5	3.428\\
6	3.744\\
7	4.011\\
8	4.204\\
9	4.507\\
10	4.683\\
11	4.865\\
12	4.967\\
13	4.991\\
14	5.222\\
15	5.146\\
};
%\addlegendentry{N=8 numeric}

\addplot [color=mycolor1, forget plot]
  table[row sep=crcr]{%
1	1\\
2	1.72272058742172\\
3	2.2694456085272\\
4	2.69748430994326\\
5	3.04170020106748\\
6	3.32451990791966\\
7	3.56102485561919\\
8	3.76173081647637\\
9	3.93419407048356\\
10	4.08398409440548\\
11	4.21529628319936\\
12	4.33135109209643\\
13	4.43466168687725\\
14	4.52721800977192\\
15	4.61061620732381\\
};

\addplot [color=mycolor2, only marks, mark=o, mark options={solid, mycolor2}, forget plot]
  table[row sep=crcr]{%
1	1\\
2	1.902\\
3	2.714\\
4	3.411\\
5	4.049\\
6	4.68\\
7	5.131\\
8	5.668\\
9	6.095\\
10	6.579\\
11	6.857\\
12	7.213\\
13	7.453\\
14	7.658\\
15	7.994\\
};
%\addlegendentry{N=16 numeric}

\addplot [color=mycolor2, dashed, forget plot]
  table[row sep=crcr]{%
1	1\\
2	1.84532653268005\\
3	2.56928935265815\\
4	3.19627424532955\\
5	3.74454393967744\\
6	4.22804640357807\\
7	4.65761803649045\\
8	5.04180551093352\\
9	5.38744020133988\\
10	5.70004794514082\\
11	5.98414669649732\\
12	6.24346629577029\\
13	6.48111313697884\\
14	6.69969520484201\\
15	6.90141817974299\\
};

\addplot [color=mycolor3, only marks, mark=triangle, mark options={solid, mycolor3}, forget plot]
  table[row sep=crcr]{%
1	1\\
2	1.949\\
3	2.842\\
4	3.694\\
5	4.455\\
6	5.249\\
7	6.021\\
8	6.68\\
9	7.326\\
10	7.927\\
11	8.522\\
12	9.077\\
13	9.648\\
14	10.061\\
15	10.54\\
};
%\addlegendentry{N=32 numeric}

\addplot [color=mycolor3, dashdotted, forget plot]
  table[row sep=crcr]{%
1	1\\
2	1.9176045149044\\
3	2.76259413060009\\
4	3.54326037258931\\
5	4.26667869591551\\
6	4.93892354245237\\
7	5.56523931880115\\
8	6.15017748717758\\
9	6.69770735435372\\
10	7.21130626350455\\
11	7.69403352216988\\
12	8.14859138808934\\
13	8.57737568128987\\
14	8.98251802435222\\
15	9.36592128310637\\
};

\addplot [color=mycolor1, mark=x, mark options={solid, mycolor1}]{
 -1 -1};
\addlegendentry{N=8}

\addplot [color=mycolor2, dashed, mark=o, mark options={solid, mycolor2}]{
 -1 -1};
\addlegendentry{N=16}

\addplot [color=mycolor3, dashdotted, mark=triangle, mark options={solid, mycolor3}]{
 -1 -1};
\addlegendentry{N=32}

\end{axis}

\end{tikzpicture}%

%% file: figures/legend_SNR.tex
\begin{tikzpicture}

\begin{axis}[
    width=0,
    height=0,
    at={(0,0)},
    scale only axis,
    xmin=0,
    xmax=0,
    xtick={},
    ymin=0,
    ymax=0,
    ytick={},
    axis background/.style={fill=white},
    legend style={legend cell align=center, align=center, draw=white!15!black, font=\scriptsize, at={(0, 0)}, anchor=center, /tikz/every even column/.append style={column sep=2em}},
    legend columns=10,
]
\addplot [color=mycolor1, only marks, mark=x, mark options={solid, mycolor1}]
  table{%
 0 1
};
\addlegendentry{MRC numerical}

\addplot [color=mycolor1]
  table{%
 0 1
};
\addlegendentry{MRC theory}

\addplot [color=mycolor2, only marks, mark=o, mark options={solid, mycolor2}]
  table{%
 0 1
};
\addlegendentry{Single beam numerical}

\addplot [color=mycolor2, dashed]
  table{%
  0 1
};
\addlegendentry{Single beam theory}

\end{axis}

\end{tikzpicture}

%% file: figures/SNR_8ant_180deg.tex
\begin{tikzpicture}

\begin{axis}[%
width=\fwidth,
height=\fheight,
scale only axis,
xmin=1,
xmax=20,
style={font=\footnotesize\color{white!15!black}},
xlabel={$M$},
xlabel near ticks,
ymin=5,
ymax=25,
ylabel style={font=\footnotesize\color{white!15!black}},
ylabel={SNR [dB]},
ylabel near ticks,
axis background/.style={fill=white},
legend style={at={(0.98,0.02)}, anchor=south east, legend cell align=left, align=left, draw=white!15!black}]
\addplot [color=mycolor1, only marks, mark=x, mark options={solid, mycolor1}]
  table[row sep=crcr]{%
1	9.15332896118138\\
2	10.4265890576573\\
3	11.3914643082169\\
4	12.0443820110473\\
5	12.3675723164033\\
6	12.8267993295224\\
7	13.0116286253256\\
8	13.3275898978791\\
9	13.736162782486\\
10	13.8674839005024\\
11	14.0690799677364\\
12	14.3054457648469\\
13	14.7107127517169\\
14	14.8245879285008\\
15	14.9654679621392\\
16	15.2397580779121\\
17	15.3600069643201\\
18	15.5306400905129\\
19	15.7031500829028\\
20	15.8607719343697\\
};

\addplot [color=mycolor1]
  table[row sep=crcr]{%
1	12.0411998265592\\
2	12.3888594776924\\
3	12.7107393870259\\
4	13.0104002861335\\
5	13.2907126041471\\
6	13.5540241224368\\
7	13.8022796316512\\
8	14.0371081913019\\
9	14.2598879793924\\
10	14.4717953023562\\
11	14.6738421925294\\
12	14.8669056407577\\
13	15.0517506025874\\
14	15.229048304666\\
15	15.3993909583787\\
16	15.563303694988\\
17	15.7212543290523\\
18	15.8736614077191\\
19	16.0209008948075\\
20	16.1633117584455\\
};

\addplot [color=mycolor2, only marks, mark=o, mark options={solid, mycolor2}]
  table[row sep=crcr]{%
1	9.15332896118138\\
2	10.8705991346982\\
3	12.1576018472294\\
4	12.9446330796804\\
5	13.4008855780202\\
6	13.8865183545988\\
7	14.1878528695996\\
8	14.5239159335116\\
9	14.9749311857197\\
10	15.1551047626345\\
11	15.3102579983797\\
12	15.6017619864347\\
13	15.9933590202419\\
14	16.0540790827218\\
15	16.2199036163151\\
16	16.4644986183209\\
17	16.4956608853788\\
18	16.778174586467\\
19	16.8326886839803\\
20	17.0289255957828\\
};

\addplot [color=mycolor2, dashed]
  table[row sep=crcr]{%
1	6.64428095775111\\
2	10.6056149241413\\
3	12.0611311113355\\
4	12.946545026607\\
5	13.5845222543624\\
6	14.0856620497123\\
7	14.5003237812779\\
8	14.8554638300514\\
9	15.1671276659695\\
10	15.4456152174477\\
11	15.6979232768723\\
12	15.9290189363739\\
13	16.1425550782675\\
14	16.3412969009762\\
15	16.5273887248526\\
16	16.7025276921231\\
17	16.8680807306377\\
18	17.0251656195638\\
19	17.1747085962858\\
20	17.3174861972687\\
};

\end{axis}

\end{tikzpicture}%

%% file: figures/SNR_16ant_180deg.tex
\begin{tikzpicture}

\begin{axis}[%
width=\fwidth,
height=\fheight,
scale only axis,
xmin=1,
xmax=20,
style={font=\footnotesize\color{white!15!black}},
xlabel={$M$},
xlabel near ticks,
ymin=5,
ymax=25,
ylabel style={font=\footnotesize\color{white!15!black}},
axis background/.style={fill=white},
legend style={at={(0.98,0.02)}, anchor=south east, legend cell align=left, align=left, draw=white!15!black}]
\addplot [color=mycolor1, only marks, mark=x, mark options={solid, mycolor1}]
  table[row sep=crcr]{%
1	11.9656517970602\\
2	13.5433854861474\\
3	13.9207352211443\\
4	14.4462110218403\\
5	14.8920402963154\\
6	15.3210985044127\\
7	15.4336745907766\\
8	15.6879024271723\\
9	15.8795775338107\\
10	16.0883808006921\\
11	16.3589759497847\\
12	16.4925710255165\\
13	16.6103997635962\\
14	16.8363032432024\\
15	16.9125984066737\\
16	17.0906269805272\\
17	17.1414807188702\\
18	17.3872289853118\\
19	17.5004889077079\\
20	17.5717161036105\\
};

\addplot [color=mycolor1]
  table[row sep=crcr]{%
1	15.0514997831991\\
2	15.2464576775925\\
3	15.4330384803083\\
4	15.6119325357733\\
5	15.7837482192882\\
6	15.9490244214497\\
7	16.1082407426775\\
8	16.2618258840754\\
9	16.410164604513\\
10	16.5536035282323\\
11	16.6924560236132\\
12	16.827006325851\\
13	16.9575130399388\\
14	17.0842121324799\\
15	17.2073194993061\\
16	17.3270331790917\\
17	17.4435352699617\\
18	17.5569935956696\\
19	17.6675631596105\\
20	17.7753874182881\\
};

\addplot [color=mycolor2, only marks, mark=o, mark options={solid, mycolor2}]
  table[row sep=crcr]{%
1	11.9656517970602\\
2	14.025212667446\\
3	14.7290518824998\\
4	15.4172339003304\\
5	16.0768738365771\\
6	16.6331538719191\\
7	16.7142088614862\\
8	17.1156200682697\\
9	17.3307702147165\\
10	17.5987677161658\\
11	17.8193824445917\\
12	17.9639132828236\\
13	18.1539108279822\\
14	18.3648690626708\\
15	18.3994207697301\\
16	18.6527055509693\\
17	18.712344048935\\
18	18.9215757331272\\
19	19.1060981978085\\
20	19.1353601913835\\
};

\addplot [color=mycolor2, dashed]
  table[row sep=crcr]{%
1	9.65458091439092\\
2	13.3835765308211\\
3	14.7351199731123\\
4	15.5417239387324\\
5	16.1133441011387\\
6	16.556163776274\\
7	16.9183633692644\\
8	17.2256112709628\\
9	17.4931041089477\\
10	17.7305406759244\\
11	17.9444699263333\\
12	18.139512596584\\
13	18.3190459061371\\
14	18.4856106336325\\
15	18.6411650255765\\
16	18.7872496011877\\
17	18.9250977786677\\
18	19.0557123008876\\
19	19.1799193654546\\
20	19.298407807905\\
};

\end{axis}

\end{tikzpicture}%

%% file: figures/SNR_32ant_180deg.tex
\begin{tikzpicture}

\begin{axis}[%
width=\fwidth,
height=\fheight,
scale only axis,
xmin=1,
xmax=20,
style={font=\footnotesize\color{white!15!black}},
xlabel={$M$},
xlabel near ticks,
ymin=5,
ymax=25,
ylabel style={font=\footnotesize\color{white!15!black}},
axis background/.style={fill=white},
legend style={at={(0.98,0.02)}, anchor=south east, legend cell align=left, align=left, draw=white!15!black}]
\addplot [color=mycolor1, only marks, mark=x, mark options={solid, mycolor1}]
  table[row sep=crcr]{%
1	15.1385655753245\\
2	16.3603242323699\\
3	17.16312737528\\
4	17.3315179231525\\
5	17.7829619195045\\
6	17.8213369024476\\
7	18.1060667496879\\
8	18.299331410403\\
9	18.5040681201879\\
10	18.5855935968263\\
11	18.7797144636928\\
12	18.8212253220826\\
13	19.041624305974\\
14	19.0989288077124\\
15	19.1906231135646\\
16	19.264325695758\\
17	19.453881687043\\
18	19.3554247263324\\
19	19.4882114360354\\
20	19.651707634664\\
};

\addplot [color=mycolor1]
  table[row sep=crcr]{%
1	18.0617997398389\\
2	18.1697980987762\\
3	18.2751758406165\\
4	18.3780571385421\\
5	18.4785575440221\\
6	18.576784766966\\
7	18.6728393695821\\
8	18.76681538515\\
9	18.8588008712506\\
10	18.9488784056112\\
11	19.0371255315657\\
12	19.1236151591491\\
13	19.2084159270271\\
14	19.2915925297613\\
15	19.3732060143194\\
16	19.4533140492347\\
17	19.5319711693878\\
18	19.609228999013\\
19	19.6851364552129\\
20	19.7597399339909\\
};

\addplot [color=mycolor2, only marks, mark=o, mark options={solid, mycolor2}]
  table[row sep=crcr]{%
1	15.1385655753245\\
2	16.8288140840482\\
3	17.9896487071969\\
4	18.3937343659748\\
5	18.9990519720244\\
6	19.1951931190904\\
7	19.5763936342025\\
8	19.7491016596839\\
9	20.0844070818329\\
10	20.2502829145758\\
11	20.439092956163\\
12	20.5305073789628\\
13	20.7860621901361\\
14	20.7970944489403\\
15	20.9164878747242\\
16	21.1041382605666\\
17	21.2466397623945\\
18	21.1868219637058\\
19	21.4276885113642\\
20	21.5265068099551\\
};

\addplot [color=mycolor2, dashed]
  table[row sep=crcr]{%
1	12.6648808710307\\
2	16.2596121962702\\
3	17.5472624558947\\
4	18.3037531663705\\
5	18.8320032588877\\
6	19.2357860675811\\
7	19.5621110811362\\
8	19.835963627293\\
9	20.0720947517452\\
10	20.2798900188685\\
11	20.465666167536\\
12	20.633864828282\\
13	20.7877208009951\\
14	20.9296588493632\\
15	21.0615408487742\\
16	21.1848259641071\\
17	21.3006780023874\\
18	21.4100394567171\\
19	21.5136838616008\\
20	21.6122536261697\\
};

\end{axis}

\end{tikzpicture}%